\documentclass{aa}  
\usepackage{graphicx}
\usepackage[usenames, dvipsnames]{xcolor}
\usepackage{soul} 
\usepackage{ulem}
\usepackage{txfonts}
\usepackage[colorlinks=true,
    linkcolor=blue,
    citecolor=blue,
    filecolor=magenta,      
    urlcolor=cyan,
    breaklinks=true,]{hyperref}
\makeatletter
\renewcommand*\aa@pageof{, page \thepage{} of \pageref*{LastPage}}
\makeatother

\newcommand{\msun}{${\rm M}_\odot$}
\newcommand{\Msun}{{\rm M}_\odot}

\newcommand{\lsim}{\mathrel{\rlap{\lower 3pt \hbox{$\sim$}} \raise 2.0pt \hbox{$<$}}}
\newcommand{\gsim}{\mathrel{\rlap{\lower 3pt \hbox{$\sim$}} \raise 2.0pt \hbox{$>$}}}

\newcommand{\pam}{PA$_{\rm mm}$}
\newcommand{\paj}{PA$_{\rm j}$}

\newcommand{\Deltapa}{\ensuremath{\Delta{\rm PA}}}
\newcommand{\Thetamax}{\ensuremath{\Theta_{\rm max}}}

\begin{document} 

   \title{Partial alignment between jets and megamasers: Coherent versus selective accretion }
   \titlerunning{Partial alignment between jets and megamasers}
   \authorrunning{M. Dotti et al.}

   \author{Massimo Dotti
          \inst{1,}
          \inst{2,} \inst{3}\fnmsep\thanks{massimo.dotti@unimib.it}
          \and
          Riccardo Buscicchio          
          \inst{1,2,3}
          \and
          Francesco Bollati
          \inst{4}
          \and 
          Roberto Decarli
          \inst{5}
          \and
          Walter Del Pozzo
          \inst{6,7}
          \and 
          Alessia Franchini\inst{2,8}
    }

   \institute{
            Universit\`a degli Studi di Milano-Bicocca, Piazza della Scienza 3, 20126 Milano, Italy
        \and
            INFN, Sezione di Milano-Bicocca, Piazza della Scienza 3, I-20126 Milano, Italy
        \and
            INAF - Osservatorio Astronomico di Brera, via Brera 20, I-20121 Milano, Italy
        \and
            Leibniz-Institute for Astrophysics Potsdam (AIP), An der Sternwarte 16, 14482 Potsdam, Germany
        \and 
            INAF – Osservatorio di Astrofisica e Scienza dello Spazio di Bologna, Via Gobetti 93/3, I-40129 Bologna, Italy
        \and 
            Dipartimento di Fisica ``E. Fermi'', Università di Pisa, I-56127 Pisa, Italy
        \and
            Istituto Nazionale di Fisica Nucleare, Sezione di Pisa, I-56127 Pisa, Italy
        \and
            Institut für Astrophysik, 
            Universität Zürich,
            Winterthurerstrasse 190, CH-8057 Zürich, Switzerland
    }

   \date{Received 25 March 2024; accepted 06 November 2024}

\abstract{
Spins play a crucial role in the appearance, evolution, and occupation fraction of massive black holes (MBHs). To date, observational estimates of MBH spins are scarce, and the assumptions commonly made in such estimates have  recently been questioned. Similarly, theoretical models for MBH spin evolution, while reproducing the few observational constraints, are based on possibly oversimplified assumptions. New independent constraints on MBH spins are therefore of primary importance. We present a rigorous statistical analysis of the relative orientation of radio jets and megamaser disks in ten  low-redshift galaxies. We find a strong preference for (partial) alignment between jets and megamaser that can be attributed to two different causes:  coherent accretion and selective accretion. In the first case the partial alignment is due to an anisotropy in the gas reservoir fueling the growth of MBHs. In the second case the spin-dependent anisotropic feedback allows  long-lived accretion only if the orbits of the gas inflows are almost aligned to the MBH equatorial plane. A discussion of the implications of the two accretion scenarios regarding the evolution of MBHs is presented, together with an outlook on future observational tests aiming at discriminating between the two scenarios and   checking whether either   applies to different redshifts and black hole mass regimes.  
}

   \keywords{Accretion, accretion disks -- Black hole physics --  Methods: statistical  --  quasars:
supermassive black holes 
}

   \maketitle
%
\section{Introduction}

Astrophysical black holes (BHs) are described by their masses and their spin 
vectors only.
The latter, more than mass, can inform us  on the physical processes responsible for their fueling and feedback  \citep[e.g.,][]{BZ77,BertiVolonteri08}. 
Spin evolution through gas accretion is particularly relevant for massive BHs \citep[MBHs, with masses $M_{\rm BH}\gsim 10^5$ \msun; e.g.,][]{1971ApJ...170..223C, Soltan82, Merloni08, Shankar13} as the main process shaping the MBH spin distribution. Only the heaviest MBHs, where gas-poor binary mergers may be relevant, might be an exception \citep[e.g.,][]{fanidakis11, barausse12}. 

Until a decade ago only two models for accretion-driven spin evolution were available in the literature. The first, called coherent accretion, assumes that MBHs accrete at least their initial mass from disks with fixed orientation\footnote{Under this assumption the orientation of the disks might be different for different accretion episodes without altering the predictions of the model.} during each accretion episode. 
Under such conditions the MBH spin rapidly aligns with its accretion disk, and the following accretion spins the MBH up to values of the adimensional spin parameter $a \sim$ 1 \citep{Bardeen70, BP75}. 
The alternative chaotic accretion model assumes small and isotropically distributed accretion episodes, during each of which the BH increases its mass by less than 10$\%$, leading to slowly rotating MBHs \citep[with $a<0.2$,][]{KingPringle06}.

The time evolution of the accretion disk orientation might  differ, however,  depending on the fueling mechanisms \citep[e.g., minor and major galaxy mergers, or bar-within-bar instabilities~][]{Combes03}. 
\cite{Dotti13} first observed that the distribution of $a$ does not need to be so dichotomic. 
By relaxing the assumption of perfect coherence or isotropy, \cite{Dotti13} predicted that MBHs with masses < $10^7$\msun should have high ($a \sim 0.9$, but not maximal) spins, while for the most massive MBHs, the spin should depend on the host-galaxy dynamics. The inclusion of the \cite{Dotti13} model in semi-analytical codes allowed,  for the first time and without fine-tuning the free parameters, us to match the distribution of $a$ derived through X-ray spectral fitting(\citealt{Sesana14}, \citealt{David20}).
It should be noted,  however, that \cite{Sesana14} somewhat arbitrarily assumed that the angular momentum direction of the accretion episodes at  subparsec scales is dictated by the large-scale ($\gtrsim 100 {\rm pc}$) dynamics  of the host's gas and stars. 
Furthermore, the spin magnitude estimates through the fitting of the X-ray Fe K emission line could be biased by assumptions of the spectral analysis \citep[e.g.,][]{Parker22}. 
For these reasons independent constraints on the small-scale orientations of accretion episodes (with respect to the MBH spin) are highly valuable. 

\cite{greene13} presented observational evidence for the close alignment between nuclear jets and the angular momentum of the gas inflow up to $\sim$1 pc, traced by megamaser disks (see \citealt{Kamali19} for a confirmation of this result). These are geometrically thin disks on close-to-Keplerian orbits around MBHs, in which the observed spectral lines (associated with molecular transitions) are produced through stimulated emission. A peculiarity of megamasers is that in order to trigger enough transitions and produce the observed luminosities, the photons must travel an extended region of the molecular disk. As a consequence, megamasers are only observed within a few degrees from edge-on, so that measuring their projection on the plane of the sky  strongly constrains their three-dimensional orientation.
Under the assumption that jets and MBH spins are aligned, such  observational evidence would imply a  substantial  alignment between the  parsec and horizon scales. 
A significantly lesser degree of alignment is present between the small-scale (horizon/parsec scales, traced by jets and megamasers respectively) and the large-scale (circumnuclear and galactic disks) structures, to the point that the planes of these structures are consistent with being independently extracted from an isotropic distribution \citep{greene13}. 
\cite{greene13} extensively discuss the possible reasons for the weak (or lack of) alignment between megamasers and galactic-scale structures. 
However, the physical processes responsible for the (partial) alignment between jets and megamaser disks (horizon and parsec scales) have not been investigated  by \cite{greene13} or by \cite{Kamali19}. To date, a straightforward explanation for such a partial alignment is missing as the relativistic effects forcing the accretion disk to align with the MBH spin \citep[i.e., the Bardeen--Petterson effect,][]{BP75} cannot operate at parsec scales (see Sect.~\ref{sec:origin}).

In this study we re-analyze the sample presented in \cite{greene13}, constraining through a Bayesian analysis the significance of the observed alignment between the jets and megamaser disks and the typical 3D degree of misalignments between the two structures (Sect.~\ref{sec:alignment}).
We then focus on the possible processes leading to such alignment, proposing two alternative scenarios, one in which the accretion process is 
responsible for the observed alignment and a new scenario  in which the alignment is required in order to sustain accretion events (Sect.~\ref{sec:origin}). 
We conclude by discussing the implications of the two scenarios and the possible ways to distinguish between them (Sect.~\ref{sec:discussion}).

\section{Bayesian evidence for jet--maser disk partial alignment}\label{sec:alignment}

\subsection{Observational sample}\label{subsec:obs-sample}

The sample we analyzed consists of ten galaxies hosting both a jet and a water megamaser (observed at 22 GHz). All the hosts of such active galactic nuclei (AGN) are disk galaxies (either spirals or S0s) and have luminosities close to $L^{*}$\footnote{Where $L^{*}$ is the scale parameter of the Schechter luminosity function, for reference of the same order as the Milky Way luminosity.}, as expected given the MBH mass estimates for these sources (in the $10^6-10^7$\msun \, range).
The redshifts and masses of the stellar component and of the central MBH for each galaxy are listed in Table~\ref{tab:sample}. The hosts do not show signs of ongoing strong interactions, and are typically found in galaxy groups, as we found cross-correlating the positions of the ten hosts with the group catalog by \citet{Tully15}.

\begin{table}
        \centering
    \caption{Main properties of the sample.}
        \label{tab:sample}
        \begin{tabular}{lccr} 
                \hline
                name & redshift & $\log (M_{\rm gal}/\Msun)$ & $\log (M_{\rm BH}/\Msun)$ \\
                \hline
        NGC 3079  & 0.004 & 10.29 &  6.38 $\pm$ 0.1$^{\rm a}$\\
        NGC 2273 & 0.006 & 10.59  & 6.88 $\pm$ 0.05$^{\rm b}$\\
        NGC 1068 & 0.004 & 10.84  &  7.24$\pm$ 0.003$^{\rm c}$\\
        NGC 2960 & 0.017 & 10.76  & 7.05 $\pm$ 0.05$^{\rm b}$\\
        UGC 3789 & 0.011 & 10.42 &  7.05 $\pm$ 0.05$^{\rm b}$\\
        NGC 1194 & 0.014 & 10.81 &  7.82 $\pm$  0.05$^{\rm b}$\\
        NGC 3393 & 0.013 & 10.74  & 7.49 $\pm$ 0.12$^{\rm b}$\\
        NGC 4388 & 0.008 & 10.8  &  6.93 $\pm$ 0.05$^{\rm b}$\\
        Circinus & 0.001 & 10.88  &  6.06 $\pm$ 0.1$^{\rm d}$\\
        NGC 4258  & 0.002 & 10.43 &  7.6 $\pm$ 0.01$^{\rm e}$\\ 
                \hline
        \end{tabular}
 \tablefoot{All redshifts are obtained from NASA/NED. The stellar masses of the hosts are obtained from colors (based on NASA/NED absolute magnitudes), following \cite{2009MNRAS.400.1181Z}. Typical uncertainties on the stellar masses are $\sim 10\%$. 
 The MBH masses are from: 
 $^{\rm a}$ \cite{2004PASJf...56..605Y}, 
 $^{\rm b}$ \cite{masses_greene10}, 
 $^{\rm c}$ \cite{2023ApJ...951..109G}, 
 $^{\rm d}$ \cite{2003ApJ...590..162G}, 
 $^{\rm e}$ \cite{2013ApJ...775...13H}.} 
\end{table}

\begin{figure}
    \centering
    \includegraphics[width=0.48\textwidth]{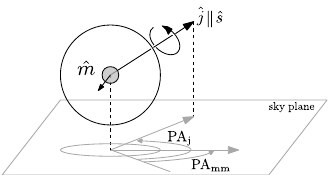}
    \caption{
     Illustration of a jet--megamaser disk system and its orientation with respect to the observer line of sight, as seen by projection angles. The jet direction (assumed to be aligned with the MBH spin $\hat{s}$) is denoted  $\hat{j}$, while the megamaser plane is defined by its normal vector $\hat{m}$. 
    Its projection onto the  sky plane is described by the angle ${\rm PA}_{\rm mm}$. The megamaser disk in the figure is close to edge-on, as in our reference model. The corresponding projection of the jet direction is denoted by the angle ${\rm PA}_{\rm j}$. For clarity, angles and vectors lying on the sky plane are shown as gray segments and arcs. }
    \label{fig:sketch}
\end{figure}

Our analysis starts from the position angles (PAs) observed for the jets (\paj) and  megamasers (\pam) of the ten AGN. The parameter 
\paj~ identifies the direction of the projection of the jet on the plane of the sky,\footnote{The PAs are quoted according  to the International Astronomical Union convention.} and it is assumed to be aligned with the direction of the MBH spin. The parameter \pam~ is the position angle of the principal axis of the ellipse obtained by projecting the megamaser disk onto the plane of the sky.\footnote{The principal axis of the projected ellipses is a good proxy for the actual 3D orientation of the megamaser disk as the stimulated emission is best traced along the disk axis, and it requires the disk plane to be nearly perpendicular to the sky plane.} An AGN with perfectly aligned jet and megamaser angular momentum will therefore result in a relative inclination between the two PAs of 90$^\circ$.
In the following we quantify the relative misalignments between megamasers and jets ($\Delta$PA) as the smallest of the two angles generated by the crossing of the two straight lines (on the plane of the sky) defined by \paj\, and \pam\, \citep[as originally defined in][]{greene13}.
A schematic illustration of the position angles is shown in Fig.~\ref{fig:sketch}.

The AGN host-galaxy names  and the measured \pam~and \paj~values are listed in Table~\ref{tab:PA}. 
Most  measurements were obtained from \cite{greene13}, though some PAs  were not quoted and were    searched for in the literature. 
Specifically, the \pam~and \paj~of NGC 3079 and Circinus are   from \cite{2005ApJ...618..618K} and \cite{2003ApJ...590..162G}, respectively; the \pam~and \paj~of NGC 1068 are   from \cite{2023ApJ...951..109G}; and the \pam~of NGC 4258 is from \cite{1995Natur.373..127M}, while its \paj~is from \cite{2018MNRAS.473.2198M}.
Since the PAs in \cite{greene13} are approximated within 5$^\circ$, we assumed each $\Delta$PA to follow a Gaussian distribution with $\sigma_{\Delta {\rm PA}}=2.5^\circ$\footnote{These uncertainties are of the same order as those quoted in the literature (when available).}. Finally, for NGC 2273 and NGC 2960 both the \pam~ and \paj~ measurements and their uncertainties are   from \cite{Kamali19}. More specifically, the uncertainty on \pam~ (\paj) for NGC 2273 is 4.6$^\circ$ (1$^\circ$), while for NGC 2960 the uncertainty of \pam~ (\paj) is 0.7$^\circ$ (2$^\circ$). The uncertainties on $\Delta$PA were then computed under the assumption that the measurements of \pam~ and \paj~ were uncorrelated, by adding in quadrature the uncertainties on the two measured angles.

\begin{table}
        \centering
    \caption{AGN host galaxy names, jet position angles, megamaser position angles, and jet--megamaser projected misalignments used in our analysis.}
        \label{tab:PA}
        \begin{tabular}{lccr} 
                \hline
                name & PA$_{\rm j}$ $[^{\circ}]$ &  PA$_{\rm mm}$  $[^{\circ}]$ & $\Delta$PA$[^{\circ}]$ \\
                \hline
        NGC 3079  & 126 &  $-$10 & 44\\
        NGC 2273   & 90  & 150 & 58\\
        NGC 1068  & 11  &  $-$50 & 61\\
        NGC 2960  & 145 &  $-$130& 70 \\
        UGC 3789   & 145 & 40   & 75\\
                NGC 1194  & 56 &  160 & 76\\
        NGC 3393  & 56  &  $-$30 & 86\\
        NGC 4388  & 24  &  110 & 86\\
        Circinus & 295  &  29 & 89\\
        NGC 4258    & $-$3  &  86 & 89\\ 
                \hline
        \end{tabular}
\end{table}

\subsection{Models}\label{subsec:pop-models}

In order to gauge whether a partial 3D alignment between the jets and megamaser angular momenta is statistically preferred over isotropic relative orientations, we built mock distributions of $\Delta$PA. 
For the isotropic case we fixed the direction of the normal to the megamaser disk ($\hat{m}$) to be the $z$ direction, and drew the direction of the MBH jet from a uniform distribution on the sphere  (i.e., with a constant probability density $p$ for the azimuthal angle and $p\propto \sin(\theta)$ for the polar angle). For each jet--megamaser realization we extracted two directions for the line of sight (LoS, $\hat{l}$), one isotropically distributed (ISO--ISO model, where the first and the second ISO refer to $\hat{j}$ and $\hat{l}$, respectively), and the other with $\theta=90^\circ$ and a random azimuth (ISO--EO model, where EO refers to the former, which is edge-on). 
The second distribution is expected to  model the observed AGN more closely as the megamaser disk is close to edge-on. For this reason we   consider as reference models those in which the megamaser is observed edge-on. However, since the $\hat{m}$ is not required to be exactly perpendicular to $\hat{l}$, the two ISO--ISO and ISO--EO scenarios bracket  the impact of  the megamaser plane inclination uncertainty. 

\begin{figure}
    \includegraphics[width=0.50\textwidth]{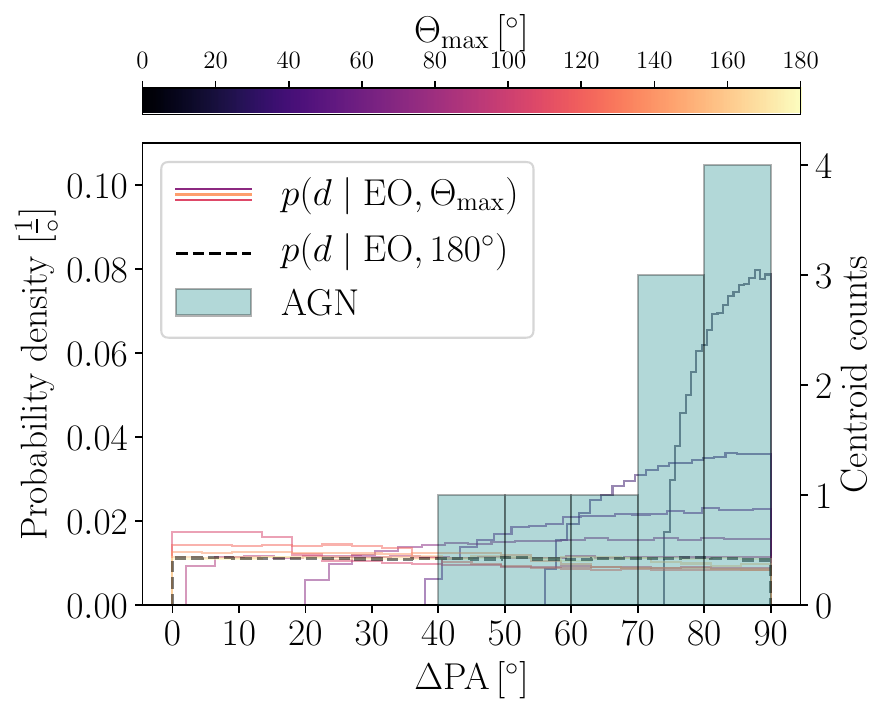}
    \centering
    \caption{Population distributions and measurements of misalignment in the plane of the sky between the megamaser disk projected semi-major axes and the jet direction ($\Delta\mathrm{PA}$). 
    The teal histogram refers to centroid counts from observed data in \cite{greene13}. The statistical analysis is performed approximating   a truncated normal distribution each $\Delta {\rm PA}$ observation, as described in Sect.~\ref{subsec:stat-analysis}.
    The colored lines denote population distributions for a few choices of $\Theta_{\rm max}$, for the edge-on model. For reference, the dashed black line denotes the $\Theta_{\rm max} = 180^\circ$ model. An absence of measurements below $\sim 40^\circ$ constrain the population posterior in Fig.~\ref{fig:samplederrors} to be above $41^\circ$ at $95\%$ confidence. Similarly, the majority of observations corresponding to $\Delta$PA $\gtrsim 70^\circ$ constrain the posterior below $68^\circ$ at $95\%$ confidence.} 
    \label{fig:histograms}
\end{figure}

In the competing scenario, we assumed the same probability distributions for $\hat{j}$ and $\hat{l}$, but limited the polar angle of the jet ($\theta_{\rm j}=\arccos(\hat{m} \cdot \hat{j})$, i.e., the angle between the  megamaser disk normal and the jet axis) to be smaller than a limiting angle $\Theta_{\rm max}$ that varies between 1$^\circ$ and 180$^\circ$ with a step of 1$^\circ$, resulting in 180 A--ISO models (in which the LoS is isotropically sampled) and 180 A--EO models.
Each model (regardless of the assumptions on the jet and LoS distributions) has been sampled with $n_{\rm s}=10^5$ jet--megamaser LoS realizations to minimize the impact of statistical fluctuations. 
In Fig.~\ref{fig:histograms} we show the  distribution of $\Delta$PA in the observed sample together with  ten mock distributions of $\Delta$PA (normalized to the number of observed systems) for equally spaced values of $\Theta_{\rm max}$ in the A--ISO case (the mock model-dependent histograms were renormalized to the number of total systems in the observed sample).  We note that   $\Theta_{\rm max}=180^\circ$ corresponds to the ISO--ISO scenario as well. A summary of the schematic models  is available in Table~\ref{tab:models}.

\subsection{Statistical analysis}\label{subsec:stat-analysis}

We now focus on connecting the predictions in Sect. \ref{subsec:pop-models} to the observations in Sect. \ref{subsec:obs-sample} through a Bayesian formalism. 
More precisely, our aim is to infer the population distribution of $\Deltapa$ from a sample of uncertain observations, described by a set of $N_{\rm AGN}$ posterior distributions obtained from the data: $\lbrace p(\Deltapa \mid d_i)\rbrace_{i=1}^{N_{\rm AGN}}$. 
For brevity, we  refer to the set of $N_{\rm AGN}$ data as $d = \lbrace d_i \rbrace_{i=1}^{N_{\rm AGN}}$.
This is a hierarchical inference problem. 
The family of models we explore is the Cartesian product of ($i$) a discrete two-element model for $\hat{m}\cdot \hat{l}$ (with equal prior probability)  parameterized by $\Psi={{\rm ISO}, {\rm EO}}$ with ($ii$) a continuous family of models for the angular distribution of the jet, $\hat{j}\cdot\hat{l}$, parameterized by $\Thetamax$. 
In particular, we denote with ``ISO'' the  model with $\cos\Thetamax=-1$ and with ``A'' any other choice of $\Thetamax$. 
The resulting family of models is summarized in Table~\ref{tab:models}.
We set priors on the A- and ISO- models as follows: 
in the partially aligned cases, the prior for $\Theta_{\rm max}$ is weighted proportionally to the solid angle within which the realizations of the jet directions are sampled. 
Including the normalization, the prior distribution reads 
\begin{equation}
\Pi_{\rm A}(\Theta_{\rm max})=\frac{1-\cos(\Theta_{\rm max})}{\pi}.   
\end{equation}
We chose this scaling simply because the possible three-dimensional configurations of the jet scale as the solid angle defined by $\Theta_{\rm max}$. Since our choice tends to favor large values of $\Theta_{\rm max}$ (compared to, e.g., a flat prior), we consider it a conservative choice when searching for a possible partial alignment between jets and megamasers.

On the other hand, models with isotropically distributed jet directions assume 
$\Theta_{\rm max}=\pi$, that is a prior $\Pi_{\rm ISO}(\Theta_\mathrm{max})=\delta (\Theta_\mathrm{max}-\pi)$, where $\delta$ denotes the Dirac delta distribution. 
The evidence of each model is  evaluated by numerically integrating the product of the prior and the likelihood over the prior support
\begin{equation}
    \mathcal{Z_{\rm mod}}=\int_0^{\pi} \mathcal{L_{\rm mod}}  \Pi_{\rm mod} \ \mathrm{d}\Theta_{\rm max} \, .
\end{equation}
For the ISO--ISO and ISO--EO models (i.e., for the models in which $\Theta_{\rm max}$ has a fixed value of $180^\circ$) the evidence corresponds to the likelihood of the model, which, in turn, is equal to the likelihood of the partially aligned case with $\Theta_{\rm max}=\pi$.

As discussed in Sect.~\ref{subsec:obs-sample}, we approximate the posteriors on $\Deltapa$ from individual and independent data $d_i$ as univariate normal distributions of known mean and fixed standard deviations, 
\begin{align}
 \Deltapa \mid d_i &\sim \cal N(\mu_{\rm \Delta PA}, \sigma_{\rm \Delta PA}), \,
\end{align}
suitably truncated and normalized over the physical range $\left[0^\circ, 90^\circ\right]$.
We connect the likelihood of a given population model (conditioned on given values of $\Thetamax,\Psi$) to the independent uncertain observations using Bayes' theorem and the probability chain rule as follows:
\begin{align}
    \mathcal{L_{\rm mod}} &\left(\left\{d_i\right\}_{i=1}^{N_{\rm AGN}} \mid \Theta_{\rm max}, \Psi\right) 
    = \!\!\prod_{i=1}^{N_{\rm AGN}}\!\!\int\!\! \mathrm{d}\Delta{\rm PA}_i p\left(d_i,\Delta{\rm PA}_i \mid \Theta_{\rm max},\!\! \Psi\right) \\
    \!\!= &\prod_{i=1}^{N_{\rm AGN}}\!\!\int \!\!\mathrm{d}\Delta{\rm PA}_i 
     p\left(d_i \mid \Delta{\rm PA}_i, \Theta_{\rm max},\!\Psi\right) p\left(\Delta{\rm PA}_i \mid\! \Theta_{\rm max}, \Psi\right)  \\
    = &\prod_{i=1}^{N_{\rm AGN}}\!\!\int \!\!\mathrm{d}\Delta{\rm PA}_i 
     p\left(d_i \mid \Delta{\rm PA}_i\right) p\left(\Delta{\rm PA}_i \mid \Theta_{\rm max}, \Psi\right)\label{eq:pop-likelihood-1} & \\
    = &\prod_{i=1}^{N_{\rm AGN}}\!\!\int \!\!\mathrm{d}\Delta{\rm PA}_i
     \frac{p(d_i) p\left(\Delta{\rm PA}_i \mid d_i\right)}{p(\Delta{\rm PA}_i)} p\left(\Delta{\rm PA}_i \mid \Theta_{\rm max}, \Psi\right) \label{eq:pop-likelihood-2}.
\end{align}

Individual likelihoods are completely specified  by the values of $\Deltapa$, so we drop the conditions $\Thetamax, \Psi$ in Eq.~\ref{eq:pop-likelihood-1}. 
The formalism in Eq.~\ref{eq:pop-likelihood-2} allows for distinct priors on individual measurements, though in this work we assume identical priors across all measurements. 
In the Bayesian context, the knowledge acquired about individual measurement $\Delta {\rm PA}_i$, after some data $d_i$ are observed (e.g., radio maps), is often represented through posterior samples drawn from $p(\Delta {\rm PA}_i\mid d_i)$, while population models are known analytically. 
In this work we tackle the opposite scenario: measurement posteriors are approximated analytically, while population models are available only through samples:

\begin{align}
    \lbrace\Deltapa_i^{(j)}\rbrace_{j=1}^{N_s} \sim p(\Delta{\rm PA}\mid \Thetamax, \Psi) \, .
\end{align}
Therefore, for each $\Thetamax$, we approximate the $N_{\rm AGN}$ integrals of the likelihood in Eq.~\ref{eq:pop-likelihood-2} via a Monte Carlo estimation as
\begin{align}\label{eq:pop-likelihood-3}
        \mathcal{L_{\rm mod}}\left(\left\{d_i\right\}_{i=1}^{N_{\rm AGN}} \mid \Theta_{\rm max}, \Psi\right)\propto\!\!\prod_{i=1}^{N_{\rm AGN}} \frac{1}{N_{\rm s}} \sum_{j=1}^{N_{\rm s}} \frac{p\left(\Delta{\rm PA}_i^{(j)} \mid d_i\right)}{p\left(\Delta{\rm PA}_i^{(j)}\right)} \, .
\end{align}

The statistical preference of a model with respect to another is quantified by the ratio of their evidence, that is the Bayes factor $K$. 
Therefore, evaluating the likelihood in Eq.~\ref{eq:pop-likelihood-3} up to a normalization constant $\prod_{i=1}^{N_\mathrm{AGN}} p(d_i)$ suffices for our objective.

In the case of our (reference) EO models, the Bayes factor for the model in which the jets partially align with the megamaser angular momentum (A--EO) and the model in which they do not (ISO--EO) is
\begin{equation}\label{eq:KEO}
K_{\rm EO}=\frac{\mathcal{Z_{\rm A-EO}}}{\mathcal{Z_{\rm ISO-EO}}}\approx 19.1,
\end{equation}
indicating a strong preference for the partially aligned model. The preference remains strong even in the case in which the LoS is sampled independently of the orientation of the megamaser disk (ISO--ISO and A--ISO models), yielding a Bayes factor of $K_{\rm ISO}={\mathcal{Z_{\rm A-ISO}}}/{\mathcal{Z_{\rm ISO-ISO}}}\approx$ 12.6.

\begin{figure*}
    \centering
    \includegraphics[width=1\textwidth]{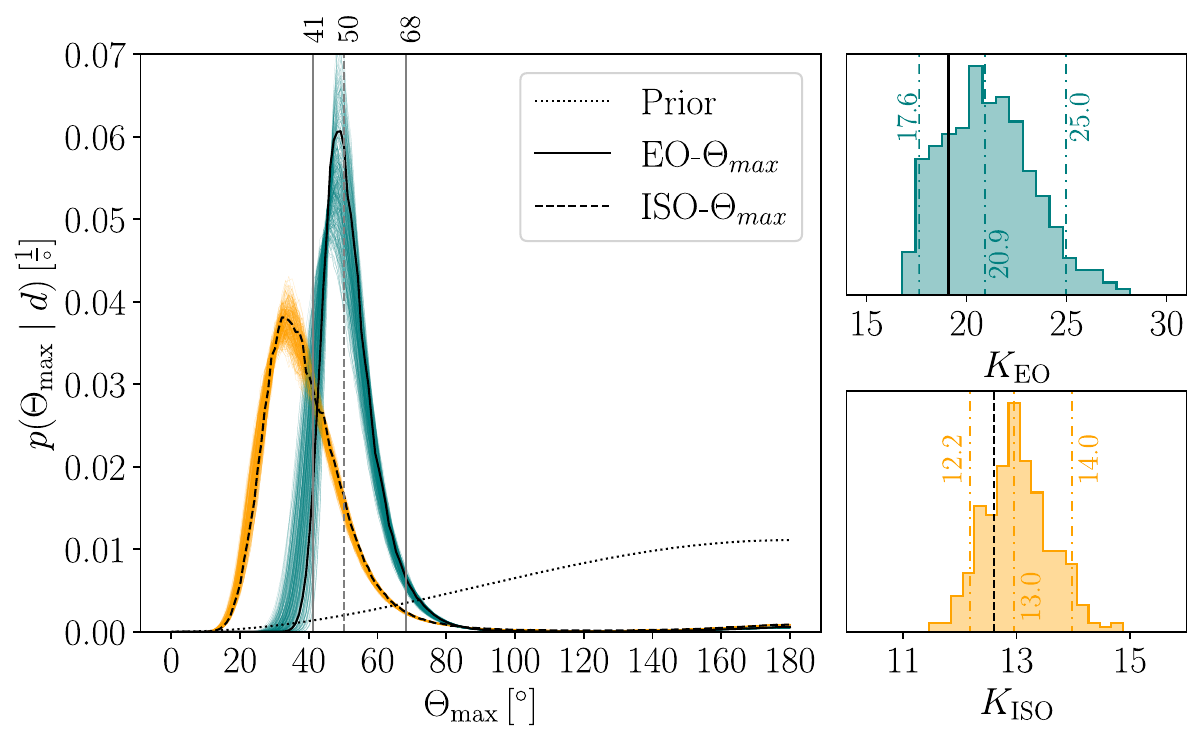}
    \caption{
    Posteriors and Bayes factors on the partial alignment angle models.
    (\textit{Left panel}) Population posteriors for the models presented in Sect.~\ref{subsec:pop-models}. The  black solid (black dashed) line denotes the distribution of the partial alignment angle $\Theta_{\rm max}$, obtained by applying the hierarchical inference scheme  in Sect. ~\ref{subsec:stat-analysis} to the EO (ISO) model. 
    The vertical dashed gray line (solid  gray lines) denotes the posterior median ($90\%$ credible interval) on $\Theta_{\rm max}$.
    The dotted line shows the prior distribution assumed for $\Theta_{\rm max}$, which has a small effect on the high-$\Theta_{\rm max}$ tail of the  mean posterior, but  is negligible for its posterior median and credible interval.
    The teal (orange) solid lines are posteriors on EO-$\Theta_{\rm max}$ (ISO-$\Theta_{\rm max}$) obtained from the 300 resampled realizations of each measurement uncertainty described in Sect.~\ref{subsec:stat-analysis}. 
    (\textit{Right panels}) Bayes factors for the partially aligned vs isotropic jet distribution model. The top (bottom) subpanel correspond to the edge-on (isotropic) LoS case. 
    Histograms denote the distribution of Bayes factors obtained from resampled measurement error realizations. The teal (orange) dash-dotted lines denote the corresponding median and $90\%$ credible interval. The solid (dashed) black line in the top (bottom) subpanel denotes the Bayes factor reported in the main text, $K_{\rm EO} = 19.1$ ($K_{\rm ISO} = 12.6$), and  obtained from the original measurement errors.
    \label{fig:samplederrors} 
    }
\end{figure*}

Our analysis indicates a clear statistical preference for the partially aligned case, as concluded in \cite{greene13} as well. For this reason we further extend our analysis by deriving the posterior distribution of $\Theta_{\rm max}$ for the A--EO model:
\begin{equation}
    p_{\rm A-EO}(\Theta_{\rm max})=\frac{\mathcal{L_{\rm A-EO}}(\Theta_{\rm max})\, \Pi_{\rm A-EO}(\Theta_{\rm max})\,}{\mathcal{Z}_{\rm A-EO}}.
\end{equation}
The resulting posterior is shown in Fig.~\ref{fig:samplederrors}. 
The posterior excludes at very high confidence $\Theta_{\rm max}=180^\circ$, with a median-centered $90\%$ confidence interval $50\substack{+19 \\ -9}^{\circ}$, respectively.
For reference, the thin dotted black line in Fig.~\ref{fig:samplederrors} denotes the prior: the inference is clearly dominated by the population likelihood except for a small increase close to the upper domain boundary. 

The posterior of the A--ISO model peaks at $\Theta_{\rm max,A-ISO}\approx 35\substack{+28\\ -15}^{\circ}$. 
The higher Bayes factor for isotropically distributed $\hat{l}$ (compared to the edge-on case) and the $\Theta_{\rm max,A-EO}>\Theta_{\rm max,A-ISO}$ inequality are both expected. Models with megamasers observed edge-on exhibit $\Delta$PA necessarily smaller than the three-dimensional angle between the jet and the megamaser. Therefore, $\Delta$PA $< \Theta_{\rm max}$. A distribution of $\Delta$PA  limited within a given angle can therefore be produced by a broader (i.e., less aligned) distribution of $\hat{j}$ and $\hat{m}$, resulting in larger $\Theta_{\rm max,A-EO}$.
A partially aligned model in which $-1\leq\hat{m} \cdot \hat{l}\leq1$ (A--ISO model) can instead result in large $\Delta$PA even if $\hat{j}$ and $\hat{m}$ are closely aligned. For instance, if the  two vectors are aligned within a few degrees, but the megamaser (or, equivalently, the jet) is even more aligned with the LoS, $\Delta$PA can be arbitrarily large, and its distribution tends toward being flat for $\hat{m} \cdot \hat{l} \to 1$. 
A--ISO models can therefore result in distributions of $\Delta$PA limited within a given angle, even if $\Theta_{\rm max,A-EO}$ is smaller than the value needed in the corresponding A--EO model, resulting in smaller  $\Theta_{\rm max,A-ISO}$. 

We note that our results depend on the assumed measurements errors on \pam\ and \paj. Since most of the uncertainties were not   presented in the original observational papers (see Sect.~\ref{subsec:obs-sample}), we quantified whether underestimated or overestimated uncertainties could severely affect our findings. 
We   therefore performed the same analysis on a sample of 300 synthetic datasets, keeping the same estimates of PAs, while drawing each measurement error from a uniform distribution between one-half and two times the assumed uncertainty. 
The analysis performed on the 300 realizations yields results that are largely compatible with our original conclusions, with Bayes factors ranging at $90\%$ confidence in the intervals $K_{\rm EO}=20.9^{+4.1}_{-3.3}$ and $K_{\rm ISO}=13.0^{+1.0}_{-0.8}$.

In order to further prove the robustness of our results, we ran an alternative test in which we compared the number of AGN observed in fixed bins of $\Delta$PA ($n_{\Delta \rm PA}$) with the corresponding expected number of AGN ($\lambda_{\Delta \rm PA}$) for each model and each $\Delta$PA. The bin size was arbitrarily chosen to be $10^{\circ}$, as in \cite{greene13}.
In this case the prior is the same as in the previous analysis, while the  likelihood $\mathcal{L}$ of each model is estimated assuming a Poissonian probability distribution,
\begin{equation}
\mathcal{L_{\rm mod}} =\prod_{j=1}^9\frac{\left(\lambda_{\Delta {\rm PA},j}\right)^{n_{\Delta {\rm PA},j}}}{\left(n_{\Delta {\rm PA},j}\right)!} \exp{\left(-\lambda_{\Delta {\rm PA},j}\right)},
\end{equation}
where ``mod'' can stand for  ISO--ISO, A--ISO, ISO--EO, or A--EO, and $n_{\Delta {\rm PA},j}$ ($\lambda_{\Delta {\rm PA},j}$) denotes the bin count (distribution parameter) of the $j$-th bin.

In order to constrain the effect of the uncertainty on the measured PAs of the jets and megamasers, we re-sampled the observed data $n_{\rm bs}=4\times10^4$ times, assuming that each PA follows a Gaussian probability distribution centered at the observed value quoted in Table~\ref{tab:PA} and with a $\sigma_{\Delta {\rm PA}}$ equal to the 2.5$^{\circ}$ uncertainty previously assumed. 
Since the number of bins in the likelihood analysis is limited to nine, the chosen value of  $n_{\rm bs}$ is large enough to cover all the possible outcomes and to constrain the average Bayes factors and their uncertainties. 
The median Bayes factor for models with edge-on megamaser disks  on the resampled data is $K_{\rm EO-bs}\approx 19.7$,
and is greater than 4.7 (12.6) at 90\% (85\%) confidence.
We note that the lower bound at 90\% is driven  by NGC3079: upon resampling, it contributes to the $\Deltapa$ bins below $40^\circ$, thus reducing the evidence for partially aligned models.
Similarly the median ratio for the models with isotropically distributed LoS is $K_{\rm ISO-bs}\approx{12.6}$ (with the ratio being greater than 7.7 (8.4) at 90\% (85\%) confidence) in good agreement with the first analysis, which is not affected by the arbitrary choice of the $\Delta$PA bin size. The posterior distributions for the partially aligned cases are consistent with those obtained with the first analysis presented.

 While a partial alignment between the jets and the megamaser is statistically preferred, our analysis indicates that, once projection effects are included, the degree of alignment in three-dimensions can be lower than what is shown by two-dimensional projections of the two structures on the plane of the sky. 
 The angle between the jet and the normal to the megamaser in the plane of the sky ($=90^{\circ}-\Delta$PA) is limited to $\approx 30^{\circ}$ for all but one of the systems in our observed sample,\footnote{The only system with a larger misalignment is NGC 3079, whose observed misalignment on the plane of the sky is $\approx 45^{\circ}$.} in agreement with the analysis on a four-AGN sample (two of which were not present in the \citealt{greene13} sample) by \cite{Kamali19}, who found a maximum misalignment of $32^{\circ}$. 
The maximum three-dimensional misalignment angle $\Theta_{\rm max}$ is somewhat larger, in particular when the information about the close-to-edge-on orientation of the megamaser is taken into account ($\Theta_{\rm max,A-EO}\lsim 68^{\circ}$  within the 90$\%$ confidence interval).

\begin{table}
        \centering
    \caption{Summary of models}
        \label{tab:models}
        \begin{tabular}{c|c|c}
  & 
  \begin{tabular}{@{}c@{}}isotropic \\
  jet distribution \\
  $-1 \leq \hat{m} \cdot \hat{j}\leq 1$ \end{tabular}
  & \begin{tabular}{@{}c@{}}partial\\
  alignment\\
  $-1 \leq \hat{m} \cdot \hat{j}<\cos(\Theta_{\rm max})$ \end{tabular}
  \\
  & & \\
  \hline
  & & \\
  \begin{tabular}{@{}c@{}}isotropic LoS \\
  $-1\leq \hat{m} \cdot \hat{l}\leq1$ \end{tabular}
  & ISO--ISO 
  & A--ISO \\
  & & \\
  \hline
  & & \\
  \begin{tabular}{@{}c@{}}edge-on\\
  megamaser \\ 
  $\hat{m} \cdot \hat{l}=0$\end{tabular} 
  & ISO--EO 
  & A--EO \\
  \end{tabular}
\end{table}

\section{Possible astrophysical origins of the observed alignment}\label{sec:origin}

The observed partial alignment between the parsec-scale Keplerian disk and the MBH spin (under the assumption that the jet traces the direction of the spin) requires a physical interpretation. 

The Kerr metric of the MBH may modify the accretion disk plane, possibly inducing a warp (i.e., the direction of the disk angular momentum changes with radius inside the disk). 
The disk warping is a direct result of the frame-dragging Lense--Thirring  \citep[LT,][]{lt1918} effect on orbits whose angular momentum is misaligned with respect to the MBH spin. The different precession rates of the disk annuli coupled with the disk viscosity, which drives an inward radial drift, also causes   a twist in the disk. This results in a  nonzero torque component in the disk orbital plane, but perpendicular to the line of nodes that tends to align (or anti-align) with the MBH spin of the angular momentum of the fluid elements within the disk. This is known as the Bardeen--Petterson effect \citep{BP75}. 
There are two distinct regimes where the warp disturbance can propagate in the disk depending on the ratio of the viscosity parameter of the disk, $\alpha$, and its aspect ratio $H/R$ \citep{Papaloizou83}: the diffusive and the bending-waves regime \citep{LOP,LodatoPringle2007,Nixon2016,Martin2019}. In the former ($\alpha > H/R$), viscosity acts to dissipate the warp, therefore aligning the disk angular momentum to the spin out to a characteristic radius $R_\textrm{warp}$ before the warp can propagate throughout the disk.  
Instead, in the latter ($\alpha < H/R$) the sound speed inside the disk is high enough so that the warp propagation occurs much more rapidly than the viscous alignment. 
In this case, the warp reflects at the disk boundaries; the disk eventually reaches a 
warped configuration and starts precessing as a rigid body around the MBH spin.

Therefore, one could, in principle, naively explain the very large observed $\Delta$PAs (i.e., the close alignment between jets and megamasers) as a consequence of the disk alignment or counter alignment 
with the MBH spin.
However, in the diffusive regime, $R_{\rm warp}$ is limited by the LT precession timescale. LT precession is indeed appreciable out to $\sim 1000$ gravitational radii \citep[$\sim 5\times 10^{-5}$ pc for a $10^6$ \msun~MBH, much smaller scales than those probed by megamasers; see, e.g.,][]{Perego09},
while the timescale for LT precession becomes longer than a Hubble time at parsec scales.
In the bending-wave regime, on the other hand, the sound speed required for a solid body precession of a whole disk extending up to parsec scales would be unphysically high and, even in this case, the steady-state wave-like warp at large distances is not expected to be aligned with the MBH spin as the alignment timescale would   again be much longer than a Hubble time \citep{Bate2000}.

For the reasons above, we need to find an alternative explanation for the observed distribution of $\Delta$PA instead of relying on the Bardeen--Petterson effect.  
The first possibility is that instead of the MBH spin aligning the outer radii of the accretion disk, it is the outer accretion disk that aligns the MBH spin with its angular momentum vector. 
The disk outer parts, misaligned with the MBH spin, supply material to the inner parts through the warp radius. As the inner parts of the disk quickly align with the MBH spin,  
if the accretion proceeds with a close-to-constant direction of the (outer) angular momentum for long enough (i.e., accreting $\gsim 1\%$ of the initial MBH mass), the spin of the MBH can move toward  alignment with the global angular momentum of the accretion reservoir \citep[e.g.,][]{Perego09}.

It is important to note that if the energy of the jet is extracted from the MBH rotation energy, MBHs with higher spins are expected to be jetted more often,  and their jets to be more luminous \citep{BZ77}.
As a consequence, our sample of jetted systems might be biased toward higher spins, achieved through prolonged gas accretion with some degree of alignment of the net angular momentum (i.e.,  far from the chaotic accretion regime with a close-to-isotropic rain of small gas clouds). Such an anisotropic feeding would automatically result in a partial alignment between the megamaser at parsec scales and the MBH spin \citep{Dotti13}.
This explanation for the observed (partial) alignment has two shortcomings. 
On the one hand, it requires some degree of fine-tuning in the orientation of the fueling events, possibly (see the discussion in the next section) in tension with the reduced evidence of alignment between the megamasers (or jets) and larger scale ($\sim 0.1-10$ kpc) structures observed in the \cite{greene13} sample.\footnote{We applied our analysis to the $\Delta$PA between megamaser-circumnuclear disks presented in \cite{greene13}, confirming that there is no strong statistical evidence of partial alignment (with an average Bayes factor of a few). A larger sample is needed to allow for any deeper insight on the nuclear-to-galactic relative orientation.} 
The second shortcoming is that it is unclear if large initial misalignments between the outer accretion disk and the MBH spin would ever eventually result in partial alignment \citep[e.g.,][see the discussion below]{Ogilvie99}.

A second possibility is represented by selective accretion, which is the triggering of accretion events only when the large-scale gas inflow is aligned (or counter-aligned) with the MBH spin, while other feeding processes with an inclination of angular momentum with respect to the MBH spin direction closer to 90$^{\circ}$ are choked on timescales short enough to become irrelevant for the MBH evolution and statistically hard to observe.
Two physical processes could be responsible for  selective accretion. 
On the  one hand, the nonlinear dynamical analysis of warped viscous disks discussed by \cite{Ogilvie99} indicates that, for sufficiently large warps (i.e., large misalignment) the viscous torque component responsible for the gas inflow reverses its sign, breaking the accretion disk in independently precessing regions, potentially stopping accretion.
However, we  note that this accretion quenching is not observed in numerical simulations. 
Disk breaking was first investigated with purely hydrodynamical simulations by \cite{nixon12}, who found continuous and rapid accretion as a consequence of the angular momentum cancellation between independently precessing detached-disk annuli. 
The dynamics of a disk subject to LT precession was further   investigated with MHD simulations, where the angular momentum transport was not modeled with a simple viscosity parameterization~\citep{sorathia13b,sorathia13a,liska19}. 
They  showed that thin accretion disks, initially strongly misaligned with the MBH spins, lead to prolonged, though modulated, accretion over multiple disk breaking episodes \citep{liska21}.

Selective accretion could, however,   also be enforced by the small-scale radiative feedback self-generated by the accretion process. The angular radiation pattern emitted from the innermost regions of the accretion disk (well within the warp radius and perpendicular to the MBH spin) is always negligible in the MBH equatorial plane, while it peaks in the spin directions for small spins or at some intermediate inclinations for higher values of $a$ \citep[e.g.,][]{Campitiello18}. In a scenario where the angular momentum of the inflowing material is close to either alignment or counteralignment with the MBH spin, a small fraction of the emitted luminosity would impact on the accreting material, limiting the effect of feedback on the gas reservoir at (sub)parsec scales. If the inflowing material has some significant misalignment with the MBH  spin, as soon as an intense accretion event starts, the outer regions of the accretion disk would be hit by a strong feedback that could eject the remaining gas quenching the accretion event. 
The threshold in terms of misalignment angle and accretion rate to trigger such feedback-driven selective accretion will depend on the value of the spin parameter. In this respect, it is worth noting that the most probable value of $\Theta_{\rm max}$ ($\lsim 50^{\circ}$) found in our analysis is comparable to the inclination required to maximize the irradiation of the outer disk for an anisotropic, spin-dependent, radiation pattern emitted by the most central part of the accretion disk\footnote{Assuming that the inner accretion disk is perpendicular to the MBH spin.} for a spin magnitude of $a\approx 0.9 $ \citep{Campitiello18}. 
An in-depth analysis of this process is left for future investigation.

\section{Implications of the observed alignment and future prospects}\label{sec:discussion}

Our study quantified the statistical evidence for a significant degree of three-dimensional alignment between the direction of the jet propagation ($\hat{j}$) and that of the angular momentum of 0.1--1 pc-scale megamaser disks ($\hat{m}$). In our reference model, where the megamaser disk plane is realistically assumed to be observed edge-on, the angle between $\hat{j}$ and $\hat{m}$ is constrained to be smaller than $\Theta_{\rm max, EO}= {49^{\circ}}^{+19^\circ}_{-10^\circ}$ at $90\%$ confidence. 
Relaxing the assumption of edge-on megamaser disks, the degree of alignment is even higher (at the same confidence level $\Theta_{\rm max, ISO}={35^{\circ}}^{+28^\circ}_{-15^\circ}$). This further demonstrates the solidity of our findings.

Under the assumption that $\hat{j}$ traces the direction of the central MBH spin, our analysis opens the exciting possibility to link the properties of the accretion process at parsec scales with the proximity of the MBH horizon ($\lsim 10^{-5}$ pc, for the systems considered in this study). This link informs us about the MBH fueling process and/or the effect of accretion feedback. The partial alignment between nuclear and parsec-scale structures may be due to a preferred angular-momentum direction of the gas feeding MBHs in rotationally dominated structures \citep[as assumed by][to reproduce the observed distribution of MBH spins]{Sesana14}. In this case, the alignment has profound implications on the whole MBH population. Such ordered accretion would imply high spins (and high radiative efficiencies), in agreement with the currently 
available constraints \citep[see, e.g.,][]{brenneman13,reynolds13,reynolds21}. 
Such prediction could be independently tested  by the future Laser Interferometer Space Antenna \citep[e.g., LISA][]{lisa1,lisa2},  sensible to the masses of the MBHs  in the sample analyzed \citep[$\sim 10^7$ \msun, ][]{masses_greene10}, either through the modeling of extreme-mass-ratio inspirals or, if the same dynamics of the MBH feeding holds at higher redshift during galaxy mergers, analyzing MBH-MBH coalescences.\footnote{In the case of MBH-MBH coalescences, accretion triggered by the galaxy merger could, in principle, result in different feeding properties with respect to isolated MBHs, possibly biasing the population of binary coalescences observed by LISA \citep[e.g.,][]{Bogdanovic2007,Dotti2010}.}

It is  possible, however,  that the sample studied is not representative of the MBH population in its mass and redshift range. It should be noted that the galaxy sample we analyzed is only made up of disk galaxies, typically in galaxy groups, and does not include very small galaxies, hosting MBHs lighter than $< 10^6$ \msun,  or very massive ones, for example  massive ellipticals in galaxy clusters. Further studies are needed to test the observed partial-alignment trend against a broader range of galaxy masses and morphologies. Furthermore, as commented in the previous section, the presence of a jet (required for the estimate of the alignment degree in current analysis) could be associated with high spins that would select MBHs growing and spinning up to somewhat coherent accretion. 
In this case, the observed partial alignment would have less significant implications on the more general population of MBHs, but would provide new strong observational support to the spin-paradigm for jet formation \citep[e.g.,][and references therein]{spinparadigm95,Tcheko10}, in which jets are powered by the rotational energy of the MBH through the Blandford--Znajek effect \citep{BZ77} and the spin determines whether an AGN would be radio-loud (i.e., jetted) or radio-quiet.\footnote{MBH spins might not be the only parameter playing a role in the production of strong jets. It has been proposed that the galaxy mass and the local environment might play a role, with the most massive galaxies in rich galaxy clusters being more prone to jet emission \citep[see, e.g.,][]{radioloud19}. We recall, however, that our sample does not include massive galaxies or massive MBHs (see Table~\ref{tab:sample}).}

A third possible explanation proposed here for the first time is represented by selective accretion. In this scenario, only accretion disks that are sufficiently aligned (or anti-aligned) with respect to the MBH spin result in long-lived accretion. 
This could ($i$) contribute to the limited AGN duty cycles observed as a sizable fraction of inflow episodes could not result in an observable AGN, or  ($ii$) modify the MBH spin evolution predicted in current models, requiring the implementation of selective accretion prescriptions in models used to reproduce the observed spin distributions, as in \cite{Sesana14}. 
Finally, selective accretion could exacerbate the problem of growing the heaviest  MBH at high redshifts ($z>6$), hinting at the occurrence of significant super-Eddington accretion events \citep[e.g.,][]{MDH14, VSD15}.

The biggest limitation of our analysis is represented by the limited size of the AGN sample available.\footnote{The modeling of the jet--megamaser realizations is very simple as well, but the size of the sample of AGN used is too small to discriminate between the current model and more complex alternatives.} 
The requirement of observing both a jet and a megamaser limits the number of systems of interest to about ten \citep{greene13,Kamali19}. 
 Increasing the sample of AGN with tracers of  horizon and $> 0.1$ pc (up to kiloparsec) scales is therefore a priority. 
 Due to the enhanced sensitivity of upcoming radio facilities, such as the ngVLA, the SKA--Mid (for the 22 GHz water maser), or the DSA-2000 \citep[for hydroxyl masers; see][]{Hallinan19}, targeted searches of extremely large samples of galaxies and blind searches in areas on the order of one square degree or more (particularly profitable when approaching cosmological distances) are foreseen to provide the detection of many more water or hydroxyl megamaser sources \citep{Tarchi2020,Tarchi2023}. 
 In particular, from dozens of  new megamaser sources up to a few hundred  are expected to be detectable, both locally and up to $z\sim 3$, especially if, as indicated in the literature, the water maser luminosity function evolves with redshift.
 The expected number of detected megamasers also strongly depends on the area covered by searches, and on the width of the redshift ranges probed \citep[][and Tarchi et al.~in prep.]{NGVLA24,Tarchi2024}.  

The statistical analysis discussed in Sect.~\ref{subsec:stat-analysis} can be used to constrain the degree of alignment of other small (MBH horizon) and large-scale ($\gsim$ pc)  galactic substructure pairs.
The simplest possibility explored in the literature is to study the $\Delta$PA between jets and the host galaxies as a whole. 
Different studies reached different conclusions, spanning from a complete lack of alignment to  clear evidence of partial alignment \citep[see Sect. 4 in][for an historical overview]{Dotti13}. 
As the size and the purity of the analyzed samples increases (by removing the contaminating radio emission from star-forming regions), a clearer tendency toward partial alignment seems to emerge \citep[see, e.g.,][and references therein]{battye09,zheng2024radio}, in agreement with the theoretical predictions from cosmological simulations following the evolution of MBH spins \citep{dubois14,peirani24}. 
We note, however, that the minor axes of the host galaxies constrained by photometry alone might not be directly correlated with a preferential angular momentum of the inflows feeding the MBH growth, in particular for dynamically hot structures such as massive ellipticals \citep[see the discussions in][]{battye09, Dotti13}.

Additionally, circumnuclear  structures of $\sim0.1\text{--}1$ kpc in size  can be used as large-scale tracers, including  stellar structures \citep[as done in][who found a lack of a statistical preference for a partial alignment between megamaser and circumnuclear disks]{greene13} and including cold molecular gas disks that, being dynamically colder, can more easily and more accurately trace the angular momentum direction of the large scale gas reservoir \citep[see, e.g.,][who found a close alignment between jets and circumnuclear disks in four out of six systems]{ruffa19}.

Adding the observational constraints on the jet and galactic structure inclinations with respect to the line of sight could further improve the analysis. Evaluating the inclination of the jet is unfortunately not straightforward
\citep[see discussion in][and references therein]{boschini24}.
\cite{ruffa2020} presented the first analysis of the 3D relative orientation between jets and circumnuclear disks on a sample of six AGN. 
Interestingly, the inclusion of the inclinations of jets and circumnuclear disks with respect to the line of sight  demonstrated that the largest misalignment on the plane of the sky observed in \cite{ruffa19}\footnote{For the galaxy IC 1531.} was a consequence of the projection effect, and that all the jet--circumnuclear disk pairs have a relative misalignment $\lsim 60^\circ$, similar to the value of $\Theta_{\rm max,EO}$ found in the reference case of our investigation.

Finally, the relative three-dimensional inclination  between  horizon  and $>0.1$ pc scales can be constrained from their inclinations with respect to the line of sight only, even without information about their PA. In this case, the sample of useful AGN can be further increased. 
The subparsec inclinations could be estimated most directly from the modeling of spatially resolved broad-line regions \citep[e.g.,][in which case an estimate of the PA would also be available]{gravity18,gravity24}, or through reverberation-mapping of optical-UV broad lines \citep[e.g.,][]{BmK, peterson93}, or even through the fitting of single-epoch UV, optical, and IR spectra  \citep[e.g.,][]{storchi03,pancoast14, horne21}. 
While this last option would be affected by significantly larger uncertainties (and each AGN would set significantly weaker constraints on its horizon-to-subparsec scale relative orientation), it would increase  the number of AGN by orders of magnitude, with an estimate of the inclination of the subparsec gas distribution. 

At horizon scales one possibility is to use the inclination of the innermost accretion disk, constrained from the modeling of broad-band X-ray spectra \citep[e.g.,][]{2024arXiv240116665D}. In this case even the relative inclination dependence on the spin magnitude could be constrained for a subsample of systems \citep[but see][for a discussion of the possible difficulties in estimating both inclinations and spins]{Parker22}. Using these constraints on the innermost accretion disk inclination (independent of the jet existence), we could test whether the observed partial alignment between small and large scales is affected by a selection effect or not.

\begin{acknowledgements} 
The authors thank the anonymous referee for the insightful suggestions and the thorough review. The authors thank Paraskevi (Vivi) Tsalmantza, Margherita Giustini, Giuseppe Lodato, Andrea Merloni, Andrea Tarchi and Eric Murphy for sharing their invaluable insight.
MD acknowledge funding from MIUR under the grant PRIN 2017-MB8AEZ and support from ICSC – Centro Nazionale di Ricerca in High Performance Computing, Big Data and Quantum Computing, funded by European Union – NextGenerationEU.
RB acknowledges support through the Italian Space Agency grant \textit{Phase A activity for LISA mission, Agreement n. 2017-29-H.0}, and CINECA HPC computing support through an ISCRA initiative grant.
This study is supported by the Italian Ministry for Research and University (MUR) under Grant 'Progetto Dipartimenti di Eccellenza 2023-2027' (BiCoQ).
AF acknowledges support provided by the ``GW-learn" grant agreement CRSII5 213497.
\end{acknowledgements}

\bibliographystyle{aa} 
\bibliography{main}
\label{LastPage}

\end{document}